# Prediction of individual progression rate in Parkinson's disease using clinical measures and biomechanical measures of gait and postural stability

Vyom Raval[1,2], Kevin P. Nguyen[1], Ashley Gerald[1], Richard B. Dewey, Jr.,[1], Albert Montillo[1,2]

[1]The University of Texas Southwestern Medical Center
[2]The University of Texas at Dallas

## ABSTRACT

Parkinson's disease (PD) is a common neurological disorder characterized by gait impairment. PD has no cure, and an impediment to developing a treatment is the lack of any accepted method to predict disease progression rate. The primary aim of this study was to develop a model using clinical measures and biomechanical measures of gait and postural stability to predict an individual's PD progression over two years. Data from 160 PD subjects were utilized. Machine learning models, including XGBoost and Feed Forward Neural Networks, were developed using extensive model optimization and cross-validation. The highest performing model was a neural network that used a group of clinical measures, achieved a PPV of 71% in identifying fast progressors, and explained a large portion (37%) of the variance in an individual's progression rate on held-out test data. This demonstrates the potential to predict individual PD progression rate and enrich trials by analyzing clinical and biomechanical measures with machine learning.

## 1. INTRODUCTION

Parkinson's Disease (PD) is the second most common neurodegenerative disease after Alzheimer's disease, with 60,000 new PD diagnoses made annually resulting in a prevalence estimate by 2020 in the USA of 930,000 people.[1] PD is characterized by a progressive loss of dopaminergic neurons, resulting in resting tremor, limb stiffness, and bradykinesia, which often manifests early on with a reduction in arm swing amplitude when walking. The primary aim of this study was to develop a model using clinical and biomechanical gait and postural stability measures capable of identifying fast PD progressors with a high Positive Predictive Value (PPV). Achieving this goal will allow enrichment of future disease-modifying drug trials with fast progressors who are most likely to show detectable changes during a trial. Gait and postural stability measures were chosen as independent variables for our models because they have been previously found to be predictive of PD risk, disease severity, and PD diagnosis.[2,3] Baseline clinical measures were also investigated because of the potential they have to be predictive of PD progression rate. The most closely related study is by Latourelle et al.,[4] where the predictive power of a composite biomarker set consisting of genetic, CSF, DaTscan, clinical and demographic features was examined. In contrast, our work examines the predictive power of gait, postural stability, clinical and demographic features.

The main contributions of this study are: (1) the development of a predictive model of an individual's PD progression rate that achieves a high PPV in identifying fast progressors suitable for enrichment of clinical trials to help expedite the development of a cure, and (2) the first models that indicate gait and postural stability measures are predictive of PD progression rate.

**Table 1** Demographics for 160 PD patients in the dataset

| Demographic | Value |
| --- | --- |
| Age | 64.5 ± 9.5 |
| Men | 54% |
| Baseline MDS-UPDRS part III score | 16.0 ± 7.9 |
| 2 year MDS-UPDRS part III score | 18.2 ± 7.6 |
| On any PD medication | 84% |
| On levodopa | 63% |

## 2. MATERIALS

Data were analyzed from 160 subjects with idiopathic PD followed longitudinally for 2 years. The subjects were part of the multi-year NIH-NINDS funded Parkinson's Disease Biomarkers Program (PDBP).[5] Patient demographics are shown in **Table 1**. Disease severity was measured using the Movement Disorder Society revision of the Unified Parkinson's Disease Rating Scale (MDS-UPDRS). The MDS-UPDRS is a four-part assessment of PD severity as measured by a trained examiner. Part III of the assessment corresponds to the motor examination and involves 18 sections which are each scored on a scale from 0 (normal) to 4 (severe). Examples of sections include speech, facial expression, and gait examinations. Some sections have subsections for each hand (LH, RH) or for each upper and lower extremity (RUE, LUE, RLE, LLE). The total part III score has a range of 0 to 132. For this dataset, a trained and MDS certified examiner with eight years of prior experience conducted the assessments.

**Figure 1** Progression of MDS-UPDRS part III score across 24 months. Individual subjects plotted as grey lines. Mean and standard deviation across subjects plotted in red. Zoomed view on right shows increasing severity longitudinally.

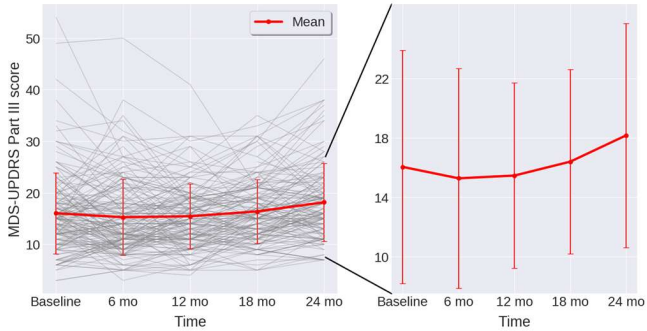

## 3. METHODS

Three targets for regression were tested individually in the following experiments: (1) total part III score at 24 months, (2) 24 months-baseline change in part III, and (3) percent change in part III measured as (24 months-baseline)/baseline. These targets were chosen as it was not known a priori whether absolute severity or a change in severity is a more predictable target. The progression of the total part III score over two years is shown in **Fig. 1**. A paired t-test revealed that the mean score at 24 months was statistically different from that at baseline ($p < 0.01$) while the mean scores at previous visits were not, indicating that 24 months is the first point at which significant progression is observed. The variability of scores across subjects exemplifies PD heterogeneity and indicates the challenging nature of the task of predicting individual progression rate.

For the gait and posture measures, subjects performed two tasks while using the APDM Mobility Lab system which uses six inertial sensors to quantify limb and torso motion. The tasks were:

1. The instrumented Timed-up-and-go (iTUG) test: subjects stand up from a chair, walk 6 meters, turn, walk back and sit down.
2. The instrumented Sway (iSway) test: subjects stand still with their feet a set distance apart and their hands across their chests for 30 seconds.

Three runs of iTUG and iSway were conducted at each visit and the median values of 148 summary statistics were utilized. Examples of measures include gait speed, step duration, stride length, arm swing velocity, and sway area. Clinical measures were used including: age, gender, baseline MDS-UPDRS part III subscores, Levodopa Equivalent Daily Dose (LEDD), and MOntreal Cognitive Assessment (MOCA) score.

**Figure 2** Feature set combinations explored. Number of features in each set indicated in parentheses. Abbreviations for feature sets in column headers.

| Feature Set | Gait and postural stability measures | | | | | Clinical | All gait and clinical |
|---|---|---|---|---|---|---|---|
| | BaseG | DeltaG | AsyBaseG | AsyDeltaG | AllG | Clin | AllGClin |
| Baseline iTUG & iSway (148) | ✓ | | | | ✓ | | ✓ |
| 6mo-Baseline iTUG & iSway (148) | | ✓ | | | ✓ | | ✓ |
| Asymmetric Baseline iTUG & iSway (22) | | | ✓ | | ✓ | | ✓ |
| Asymmetric 6mo-Baseline iTUG & iSway (22) | | | | ✓ | ✓ | | ✓ |
| Clinical measures (40) | | | | | | ✓ | ✓ |

**3.1 Feature set construction and feature selection:** Seven sets of features were compared for predictive power, encompassing different combinations of iTUG and iSway summary statistics, clinical measures, and additional derived features (**Fig. 2**). These included a set with the baseline iTUG and iSway measures (which we refer to as *BaseG*, short for baseline gait) and a set of derived features with the difference between iTUG and iSway measures at 6 months and at baseline (*DeltaG*, i.e. delta gait), which capture progression of motor symptoms. Asymmetric presentation of motor dysfunctions in PD has shown to be an important marker of PD severity,[6] so asymmetry measures on the 22 lateralized variables in baseline (*AsyBaseG*) and 6 months-baseline iTUG and iSway (*AsyDeltaG*) were computed with the formula $1 - \frac{Left\ measure}{Right\ measure}$. All of these iTUG and iSway measures were combined in feature set *AllG*, i.e. all gait measures. Clinical measures were considered by themselves (*Clin*) and in combination with all iTUG and iSway measures (*AllGClin*, i.e. all gait and clinical measures). Feature selection was conducted on the training partitions by dropping one member of each pair of highly intercorrelated features (Pearson's $r > 0.8$) to minimize feature redundancy.

**3.2 Data partitioning and model training:** XGBoost and Feed Forward Neural Network (NN) models were chosen as they are two of the most powerful models that consistently win machine-learning competitions for structured and unstructured data and have shown high performance in a wide range of tasks.[7] Mean-squared-error loss was used to train the NNs. Model performance was evaluated using the $R^2$ score, i.e., the coefficient of determination. The dataset was partitioned using nested K-fold cross validation with 3 inner and 3 outer folds. In each outer fold, mean $R^2$ across the held-out partitions of the inner folds was used to rank model performance and the model with the highest mean $R^2$ was selected for evaluation on the held-out partition of the outer fold. The mean test performance over the held-out partitions in the 3 outer folds represents final model performance. Stratified k-fold partitioning was used to ensure representative target distributions across splits and appropriate model training and evaluation.

**Table 2** Hyperparameter ranges explored for each model.

| XGBoost |
| --- |
| Number of estimators: [10, 1000] |
| Maximum depth: [5, 50] |
| L1 regularization term: (0, 1) |
| L2 regularization term: (0, 1) |
| Learning rate: [0.0001, 0.4] |

| Feed Forward Neural Network |
| --- |
| Layers: [1, 5] |
| Chance to taper: 50% |
| Taper size: {0.2, 0.5} |
| Dropout: [0.1, 1.0] |
| Activations: {ReLU, ELU, LeakyReLU, PReLU, tanh, sigmoid} |
| Number of neurons: {16, 32, 48, 64, 80, 96, 112, 128} |
| Learning rate: [0.0001, 0.005] |
| Optimizer: Nadam |

**3.3 Hyperparameter optimization and model selection:** To identify optimal model hyperparameters in an unbiased manner, a random search of 1000 hyperparameter configurations for XGBoost and 300 configurations for NNs was conducted. Fewer configurations were searched for NNs due to computational requirements. The hyperparameter dimensions and ranges searched are shown in **Table 2**. The best-performing hyperparameter configurations were selected based on mean $R^2$ across the inner cross-validation folds, and the model's performance is evaluated as the mean $R^2$ evaluated on the held-out test splits. Hyperparameter vs. performance plots [not shown] confirmed that sufficient ranges of the hyperparameters was searched such that the local maxima of performance were found.

**3.4 Feature importance:** To reveal what the NNs learned, *feature permutation importance* was used to compute feature importance. Each feature in the held-out test set was randomly permuted 100 times and the decrease in $R^2$ was measured, with a greater mean decrease reflecting greater importance.

**Figure 3** Mean test $R^2$ performances of best models on each feature-target combination. Feature sets are shown on the x-axis and targets are shown on the y-axis. The color bar indicates $R^2$, with brighter green indicating better performance and black indicating 0 or negative $R^2$ performance.

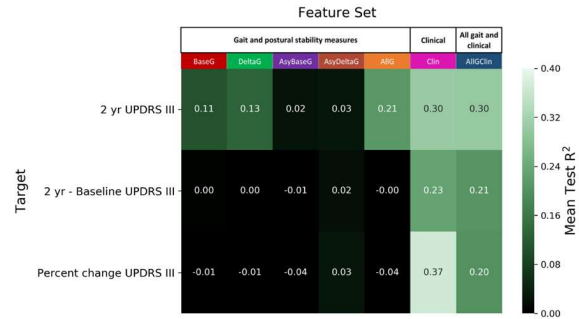

## 4. RESULTS

The mean test $R^2$ performances of the best models on each feature set × target combination are shown in **Fig. 3**. NNs outperformed XGBoost models in every case. The feature set *Clin*, which used only the clinical measures, achieved the highest $R^2$ across all prediction targets and model categories. Using this feature set, 37% of the variance was explained in the percentage change MDS-UPDRS part III score. Given that predicting progression rate is a difficult problem as indicated by the large standard deviation of ± 8 points in the progression rate across subjects, explaining nearly 40% of the variance in progression rate using our model is a significant finding. This result is comparable to the 41% validation performance achieved by Latourelle et al.,[4] and has the added strength that our evaluation is on held-out test data while theirs was on validation data which typically inflates the result. Our model achieved a PPV of 71% in identifying fast progressors, defined as having a 20% or more increase in MDS-UPDRS part III score from baseline (top 50% of the cohort).

This is also the first study to show gait and postural stability measures to be predictive of PD progression. Three of the gait and postural stability feature sets explained 10% or more of the variance in the 2 year MDS-UPDRS part III score. Feature set *AllG*, which included all the derived gait and postural stability measures, explained 21% of the variance.

The 15 most important features learned by NNs on sets *Clin* and *AllG* are shown in **Fig. 4**. The MDS-UPDRS part III RUE rigidity subscore, total score, and Right Hand Finger Tapping subscore were the three most important features for predicting percent change (**Fig. 4A**). In the *AllG* feature set, the (6 months-baseline) iTUG asymmetry values ranked high in feature importance, with the iTUG gait stride velocity asymmetry change as the most important for predicting the 2 year score (**Fig. 4B**).

**Figure 4** Mean feature importances of the top 15 features learned by the best performing NNs. **A)** Top features using feature set *Clin* to predict the percent change in MDS-UPDRS part III and **B)** Top features using feature set *AllG* to predict 2 year MDS-UPDRS part III score. Error bars represent the standard deviation of the feature importance across the outer

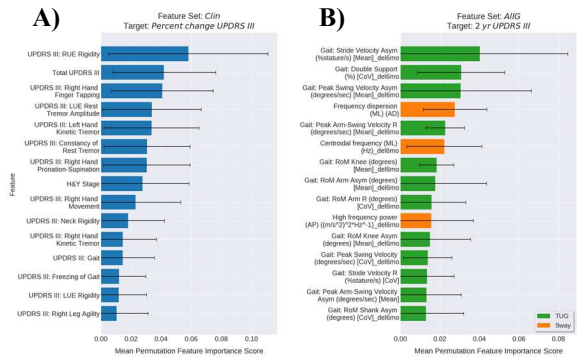

## 5. DISCUSSION

The best model performance was obtained using clinical measures to predict the 2 year percent change MDS-UPDRS part III score with an NN. The model explained 37% of the variance in the target, with a PPV of 71% in identifying fast progressors. Thus, the model may be useful in enriching disease-modifying drug trials with fast progressors. Similar to Latourelle et al.,[4] baseline movement scores from *Clin* are found to be among the most important features for predicting future progression rate.

Further, this is the first study to show that gait and postural stability measures have predictive power for PD progression. The (6 months-baseline) iTUG asymmetry values ranked high in feature importance, indicating that the progression of asymmetric aspects of gait impairments are especially important for predicting PD progression. This demonstrates the prognostic value that can be provided by improved measurements of motor disability in PD subjects. While the gait and postural stability measures alone in *AllG* performed modestly ($R^2$=0.21), their performance was bolstered by the inclusion of clinical measures in *AllGClin* ($R^2$=0.30). However, this performance boost was not additive, suggesting that the feature sets have collinearities and measure similar aspects of PD progression.

The primary limitation of this study is that a single dataset was used. Though the dataset was fairly large (N=160 subjects) and rigorous cross-validation was performed including a held-out test set not used for training or model selection, a replication study on an independent dataset would further confirm our findings. Fortunately, other datasets are becoming available and replication of our model's performance on these datasets is the subject of our ongoing research. Additional future studies are planned to increase predictive power using other methods to derive features from the iTUG and iSway sensor data. Such deeper analysis may enable even more prognostic applications.

## 6. CONCLUSION

The main contributions of this study include the development of a predictive model of an individual's PD progression rate that achieves a 71% PPV in identifying fast progressors, which is suitable to enrich clinical trials to help expedite the development of a cure for PD. This work reaffirms the importance of clinical measures in predicting PD progression and suggests the potential for gait and postural stability measures as a predictive tool.